\documentclass[12pt]{article}

\usepackage{amsmath, amssymb, amsthm}
\usepackage{fullpage}      
\usepackage{graphicx}
\usepackage{xcolor}
\usepackage{authblk}
\providecommand{\keywords}[1]{\textbf{Keywords:} #1}
\usepackage{natbib}        
\usepackage{hyperref}      
\usepackage{subcaption}
\newtheorem{thm}{Theorem}

\newcommand\independent{\protect\mathpalette{\protect\independenT}{\perp}}
\def\independenT#1#2{\mathrel{\rlap{$#1#2$}\mkern2mu{#1#2}}}

\title{Robust and Efficient Semiparametric Inference for the Stepped Wedge Design}

\author[1]{Fan Xia}
\author[2]{K. C. Gary Chan}
\author[3]{Emily Voldal}
\author[4]{Avi Kenny}
\author[2]{Patrick J. Heagerty}
\author[2]{James P. Hughes}

\affil[1]{University of California, San Francisco}
\affil[2]{University of Washington}
\affil[3]{Fred Hutchinson Cancer Center}
\affil[4]{Duke University}
\date{} 

\begin{document}

\maketitle
\begin{abstract}
    Stepped wedge designs (SWDs) are increasingly used to evaluate longitudinal cluster-level interventions but pose substantial challenges for valid inference. Because crossover times are randomized, intervention effects are intrinsically confounded with secular time trends, while heterogeneity across clusters, complex correlation structures, baseline covariate imbalances, and small numbers of clusters further complicate inference. We propose a unified semiparametric framework for estimating possibly time-varying intervention effects in SWDs. Under a semiparametric model on treatment contrast, we develop a nonstandard semiparametric efficiency theory that accommodates correlated observations within clusters, varying cluster-period sizes, and weakly dependent treatment assignments. The resulting estimator is consistent and asymptotically normal even under misspecified covariance structure and control cluster-period means, and is efficient when both are correctly specified. To enable inference with few clusters, we exploit the permutation structure of treatment assignment to propose a standard error estimator that reflects finite-sample variability, with a leave-one-out correction to reduce plug-in bias. The framework also allows seamless incorporation of effect modification and adjustment for imbalanced baseline precision variables 
    through a design-based adjustment shown to be closely related to post-stratification, or a double adjustment that additionally incorporates an outcome-based component. 
    Simulations and application to a public health trial demonstrate the robustness and efficiency of the proposed method relative to standard approaches.
\end{abstract}


\section{Introduction}
Stepped wedge designs (SWD) are a type of cluster randomized trial commonly used to evaluate interventions delivered to groups of individuals such as clinics and communities (\citealp{mdege2011systematic}; \citealp{hemming2015stepped}).  In most stepped wedge designs, all clusters start in the control condition, cross over to the intervention condition at different randomized follow-up time points and then stay on intervention until the end of the trial. 

 There are multiple challenges to generating valid statistical inference in SWDs. First, the staggered rollout of the intervention induces a structural correlation between treatment and time, making secular time trends a design-driven source of confounding. This feature distinguishes SWDs from parallel cluster randomized trials and makes appropriate adjustment for temporal trends essential. Second, heterogeneity across clusters may result in differing longitudinal time trends, violating the assumption of shared secular trends commonly adopted in parametric models. Third, with typically few clusters, randomization alone may not eliminate imbalances in baseline covariates. These imbalances can induce spurious associations between treatment and prognostic factors, leading to unstable or misleading estimates. Fourth, small numbers of clusters also limit the reliability of standard error estimation based on asymptotic approximations. Finally, the dependence structure within clusters is often complex, and mis-specification of the covariance model can further degrade the validity of statistical inference.

Conventional analysis methods in SWD are often based on mixed models (\citealp{diggle2002analysis}), which account for dependence among observations from the same cluster through a specification of random effects \citep{hussey2007design,hemming2015stepped,hooper2016sample}. However, the validity of mixed-model-based inference hinges on correct specification of both fixed and random effect components, an assumption that is often untenable in practice \citep{thompson2017bias, voldal2022misspec}. 
To mitigate concerns about misspecification of mixed models, semiparametric and nonparametric methods have been proposed. Generalized estimating equations and mixed models with robust standard errors offer protection against covariance misspecification \citep{scott2017finite, ouyang2024maintaining} while permutation-based tests leverage the randomization to provide valid inference under minimal distributional assumptions \citep{ji2017randomization, wang2017use}. However, these approaches still rely on correct specification of the outcome mean model at each time point, making them sensitive to misspecified time trends. Alternative strategies focus on within-period comparisons to avoid time-trend confounding. Notably, \cite{thompson2018robust} and \cite{kennedy2020novel} propose nonparametric estimators based on vertical comparisons within time periods, and \cite{hughes2020robust} develops a design-based estimation approach with a permutation-derived closed-form standard error estimator. While these methods improve robustness, pure within-period comparisons may be inefficient, and optimal weighting schemes remain unclear.

While randomization prevents confounding on average, small numbers of clusters can lead to sample-specific imbalances in covariates. For example, if cluster size is predictive of the outcome \citep{kahan2023informative}, an imbalance in cluster sizes will induce error in treatment effect estimates unless explicitly adjusted for. However, adjusting for such imbalances via the outcome model imposes additional assumptions and conflates design with analysis. 

In this paper, we introduce a unified semiparametric framework for effect modeling, estimation and inference in stepped wedge designs that directly addresses all of the challenges outlined above. The proposed estimator requires only the treatment contrast to be correctly specified and accommodates arbitrary user-specified treatment effect models, including, but not limited to, the commonly used instantaneous and exposure-time indicator models \citep{kennedy2020novel,billot2024should, kenny2022analysis}.  It remains robust to both misspecification of heterogeneous time trends across clusters, and the outcome’s covariance structure. The proposed estimator is consistent and asymptotically normal even when nuisance components, the control time trend and covariance structure, are misspecified, and achieves the semiparametric efficiency bound when both are correctly specified. The theoretical results are derived using a nonstandard development of semiparametric efficiency theory, in order to handle correlated observations within clusters, non-identically distributed observations across clusters due to varying cluster-period sample sizes, and weakly dependent treatment assignment across clusters that are common features of stepped wedge designs. To enable valid inference in trials with few clusters, we further develop a standard error estimator based on the permutation structure of the treatment assignment with a leave-one-out correction. We further discuss how effect modification can be naturally incorporated into the framework, while imbalanced precision variables can be addressed through either a design-based adjustment closely related to post-stratification or a double adjustment that additionally incorporates an outcome-based component.

We evaluate the proposed method through extensive simulation studies to demonstrate the performance of our proposed estimator under various types of model misspecifications and with small numbers of clusters. We specifically compare our approach with linear mixed models, the prevailing standard in SWD analysis.
Finally, we apply our proposed method to the Washington State expedited partner treatment trial data \citep{golden2015uptake}. The code for reproducing the experimental results reported in this paper is available on GitHub\footnote{\url{https://github.com/fanxiaxia/Robust-and-Efficient-Estimation-for-SWDs}}.

\section{Method}
\subsection{Semiparametric Causal Treatment Contrast Model}
Consider a stepped wedge design with $N$ clusters and $T$ measurement time points. Let $Y_{ijk}$ be the observation on individual $k$ ($k = 1,\ldots, n_{ij}$) at time $j$ ($j=1,\ldots, T$) in cluster $i$ ($i=1,\ldots, N$), and $X_{ij}$ be the binary indicator of whether the intervention is active in cluster $i$ at time $j$.  Our framework covers both cohort and repeated cross-sectional designs.

In a standard stepped wedge design with $S = T - 1$ sequences, the crossover time $R_i$ for cluster $i$ takes values from $\{2, \dots, S+1\}$, specifying the time point at which cluster $i$ transitions to the intervention. Once a cluster crosses over, it remains under the intervention for all subsequent time points.  Clusters are usually not randomized independently according to a
Bernoulli distribution; instead, they are allocated from a permutation distribution of possible assignments with a fixed number of clusters initiating the intervention at each time. Such
an assignment induces dependence across clusters \citep{hughes2020robust}.

The crossover time $R_i$ and the intervention sequence $\bar{X}_{iT} = (X_{i1}, \dots, X_{iT})$ are deterministically related. Specifically, if $R_i = r$, then $X_{ij} = 0$ for all $j < r$ and $X_{ij} = 1$ for all $j \geq r$. Thus knowing the crossover time $R_i$ fully determines  the treatment sequence $\bar{X}_{iT}$, and vice versa.

The potential outcome for an individual $k$ in cluster $i$ at time $j$, given the entire treatment history $\bar{x}_{iT}$, is denoted as $Y_{ijk}(\bar{x}_{iT})$. Because clusters are randomized to different sequences, the assumption of exchangeability holds:
\begin{align}
    ({Y}_{i1k}(r_i),\dots,Y_{iTk}(r_i))\independent R_i,
    \label{exch_R}
\end{align}
for any $i$ and $k$. This means that the potential outcomes for individuals in a given cluster are independent of the assigned treatment sequence.

Since the crossover time $r_i$ maps bijectively to a sequence of intervention indicators $\bar{x}_{iT}$, the potential outcomes can equivalently be expressed as:
\[Y_{ijk}(r_i) = Y_{ijk}(\bar{x}_{iT}).
\]
Under the assumption of no anticipation, \emph{i.e.}, future interventions do not affect past outcomes:
\[
Y_{ijk}(\bar{x}_{iT}) = Y_{ijk}(\bar{x}_{ij}),
\]
where $\bar{x}_{ij}$ represents the treatment history up to time $j$, the exchangeability condition (\ref{exch_R}) extends to:
\begin{align}
(Y_{i1k}(\bar{x}_{i1}), \dots, Y_{iTk}(\bar{x}_{iT})) \independent \bar{X}_{iT},
\label{exch_Xbar}
\end{align}
which implies that at any given time $j$, 
\begin{align*}
Y_{ijk}(\bar{x}_{ij}) \independent \bar{X}_{ij}.
\end{align*}
Therefore, at any given time $j$, confounding between the treatment sequence $\bar{X}_{ij}$ and the outcome $Y_{ijk}(\bar{x}_{ij})$ is eliminated through randomization, allowing us to study the causal effect of the treatment sequence in a vertical comparison at a given time $j$ similar to a basic randomized trial. The expected potential outcome $E[Y_{ijk}(\bar{x}_{ij})]$ is identified by $E[Y_{ijk} \mid \bar{X}_{ij} = \bar{x}_{ij}]$ given the consistency assumption $Y_{ijk}=Y_{ijk}(\bar{X}_{ij})$.  

On the other hand, pooling over time periods
typically introduces bias in a way that is analogous to confounding because 
time itself influences both the treatment distribution and the observed outcome distribution. Therefore, in stepped wedge designs, the effect of time must be thoroughly accounted for \citep{thompson2017bias, scott2017finite}, and estimation of the treatment effect in a stepped wedge design is generally model dependent. Linear mixed models of the following form are often used:
\begin{align}
    Y_{ijk} = g_\delta(\bar{X}_{ij}) + m_\gamma(j) + Z_{ij}b_{ij} +\epsilon_{ijk}, 
    \label{lmmgen}
\end{align}
where $g_\delta(\bar{x}_{ij})$ represents the treatment effect at time j with treatment history $\bar{x}_{ij}$, $m_\gamma(j)$ is the time trend for the controls, and $Z_{ij}$, which could be dependent on treatment and/or time, is the design matrix for the vector of random effects $b_{ij}$. The random effects are often assumed to be identically and independently normally distributed across $i$ and $j$, and $\epsilon_{ijk}$ is a normally distributed error term. For cohort designs, a random effect for individual $k$ can be further included. The parameter of interest $\delta$ could be a $d$ dimensional vector, depending on the specification of the treatment effect model. With a slight abuse of notation, we drop the dependence of $j$ in $g_\delta$ to indicate the treatment effect on calendar time $j$ depends on the current time only through treatment history up to that time.  This does not limit treatment effect to be dependent on a more interpretable time scale based on treatment history, such as time since treatment crossover, also known as the exposure time. 

In the instantaneous treatment effect (IT) model \citep{hussey2007design}, 
the treatment is assumed to take an instantaneous effect which does not vary over time:
\begin{align}
    g_\delta(\bar{x}_{ij}) = x_{ij}\delta \ . \label{instant}
\end{align}
The parameter $\delta$ here is a scalar that captures the instantaneous effect.

In the exposure time indicator (ETI) model of \cite{kenny2022analysis}: 
\begin{align}
    g_\delta(\bar{x}_{ij}) = \sum_{\tau=1}^{T-1}l(\bar{x}_{ij};\tau)\delta_{\tau} \ . \label{eti}
\end{align}
The function $l(\bar{x}_{ij};\tau)=\mathbf{I}(\sum_{t=1}^{j}x_{it}=\tau)$ indicates whether the exposure time of cluster $i$ at time $j$ is $\tau$, $1 \leq \tau \leq T-1$. The parameter $\delta$ here is a $T-1$ dimensional vector $(\delta_1,\dots,\delta_{T-1})$, where $\delta_{\tau}$ captures the effect of treatment after being exposed for $\tau$ periods. 
In both \cite{hussey2007design} and \cite{kenny2022analysis}, the time trend is modeled by a categorical variable (\emph{i.e.} $m_\gamma(j) = \gamma_j$, $j = 1,\dots, T$).

Both the IT and the ETI models are examples of structural models in causal inference \citep{hernan2020causal}, where we explicitly define how different variables are related to the expected potential outcomes. For these models to provide valid causal interpretations, they must correctly reflect the data-generating mechanisms. This means that both $g_\delta(\bar{X}_{ij})$ and $m_\gamma(j)$ must be properly specified. While using a saturated time model can help guard against misspecification in $m_\gamma(j)$, it is not always a satisfactory solution. In stepped wedge trials with a large number of time periods, fully flexible time models may be impractical as the number of parameters for time trends can be greater than the number of independent clusters, and it is common in practice to impose stronger assumptions, such as assuming a linear time trend to simplify estimation \citep{jarvik2020effect}.
Since $m_\gamma(j)$ is a nuisance model that is not of primary interest, we propose constructing an estimator for $\delta$ in the treatment model $g_\delta(\bar{X}_{ij})$ that does not rely on the correct specification of $m_\gamma(j)$.

Specifically, we study a semiparametric model that only specifies the causal treatment contrast:
\begin{align}
    E[Y_{ijk}(\bar{x}_{ij})-Y_{ijk}(\bar{0}_j)] = g_\delta(\bar{x}_{ij}), \label{trteffect}
\end{align}
\textcolor{black}{where $g_\delta(\bar{x}_{ij})$ satisfies $g_\delta(\bar{0}_{j}) = 0$ ($j=1,\ldots,T$).}  The treatment effect model reflects the scientific belief about the causal mechanism, where $\delta$ captures the effect of $\bar{x}_{ij}$ on the outcome through $g_\delta(\bar{x}_{ij})$ for all $j$. Note that the comparison group $E[Y_{ijk}(\bar{0}_j)]$ is not non-parametrically identifiable, since (typically) no cluster is assigned to a pure control condition.

Our model assumption is minimalist in the sense that we only impose restrictions on the form of the treatment effect, assuming a hypothesized mechanism that can lead to a reasonable model for $g_\delta$ for which the parameters $\delta$ has a clear scientific interpretation. Any other aspects of the data-generating mechanisms remain unconstrained.

Under exchangeability in (\ref{exch_Xbar}),  and the consistency assumption, we can rewrite Model (\ref{trteffect}) as
\begin{align}
    E[Y_{ijk} - g_\delta(\bar{X}_{ij}) \mid \bar{X}_{ij}] = E[Y_{ijk} - g_\delta(\bar{X}_{ij})]. \label{id_trteffect}
\end{align}
The corresponding derivation can be found in Section 1 of the Supplementary Materials.

The restriction in (\ref{id_trteffect}) implies that after removing the treatment effect $g_\delta(\bar{X}_{ij})$, the remainder of the outcome is independent of $\bar{X}_{ij}$ in mean. This restriction is implied by any structural model like (\ref{lmmgen}), including mixed models with additional random effects for time or cluster-time interactions as examined in \cite{ji2017randomization}, models with an autoregressive correlation structure as in \cite{kasza2019impact}, and models with cluster-treatment random effects since the restriction applies only to the cluster-specific mean.  Our purpose is to study valid estimators and inference for parameters $\delta$ in model (\ref{trteffect}) when the validity of additional modeling assumptions may be questionable.

\subsection{Semiparametric Theory and Estimation}
Under the semiparametric model constraint (\ref{id_trteffect}), we derive the semiparametric efficiency bound for regular estimators for $\delta$. We begin by introducing vector notations.
For a stepped wedge trial with $T$ time points, let the vector of outcomes for cluster $i$ at time $j$ be $Y_{ij} = (Y_{ij1}, \dots, Y_{ij{n}_{ij}})^T$, and the vector of outcomes for cluster $i$ across all time points be $Y_i = (Y_{i1}, \dots, Y_{iT})^T$. Denote $\mathbf{1}_{{n}_{ij}} = (1,\dots,1)^T \in R^{n_{ij}}$, the vector of treatments for cluster $i$ is $\underline{X}_i = ({X}_{i1} \mathbf{1}^T_{{n}_{i1}}, \dots, {X}_{iT} \mathbf{1}^T_{{n}_{iT}})^T$, and the treatment effect model vector is $g_\delta(\underline{x}_i) = (g_\delta(\bar{x}_{i1}) \mathbf{1}^T_{{n}_{i1}}, \dots, g_\delta(\bar{x}_{iT}) \mathbf{1}^T_{{n}_{iT}})^T$. Similarly, the time trend vector is $m_i(T) = (m_{\gamma,i}(1) \mathbf{1}^T_{{n}_{i1}}, \dots, m_{\gamma,i}(T) \mathbf{1}^T_{{n}_{iT}})^T$, where we allow the time trend to vary across clusters. Denote the derivative of $g_\delta(\underline{X}_i)$ with respect to $\delta$ as $\Dot{g}_\delta(\underline{X}_i)$, which is a matrix of size $\sum_{ij} n_{ij} \times d$.

A major theoretical complication is that standard techniques for showing semiparametric efficiency bounds and large sample results, which often assume independent and identically distributed data, cannot be applied.  Indeed, we have both dependence of treatment assignment across clusters due to restricted randomization, and non-identically distributed outcomes across clusters (e.g. cluster sizes vary). 
The corresponding efficiency bound is formulated along a sequence of finite-sample designs, each inducing a joint distribution for non-identically distributed clusters across all time points.
Details are given in Section 2 of the Supplementary Materials.

\begin{thm}
The efficient score function under (\ref{id_trteffect}) is
\begin{align}
    S_{\text{eff}}(X_i, Y_i; \delta) = \sum_i L_i^T W_i \{ Y_i - g_\delta (\underline{X}_i) - m_i(T) \}, \label{einf}
\end{align}
where 
\begin{align*}
   & W_i = \operatorname{Var}^{-1}\{Y_i - g_\delta (\underline{X}_i) - m_i(T) \mid\underline{X}_i\}, \\
   & m_i(T) = E[Y_{i} - g_\delta(\underline{X}_i)], \\
   & L_i = \Dot{g}_\delta(\underline{X}_i) - E^{-1}(W_i) E\left\{ W_i \Dot{g}_\delta(\underline{X}_i) \right\} \ .
\end{align*}
In the definition of $W_i$, the expectation is taken over the distribution of the outcome conditional on the treatment. In $m_i(T)$, the expectation is also taken over the distribution of the outcome but is not dependent on $\underline{X}_i$ according to \eqref{id_trteffect}.  In $L_i$, the expectations are taken over the distribution of the treatment assignments. The semiparametric bound for regular estimators of $\delta$ under (\ref{id_trteffect}) is the inverse of the variance of $S_{\text{eff}}(X_i, Y_i; \delta)$.
\end{thm}

Based on the efficient score function, we can construct the following estimating equations 
\begin{align}
    \sum_i\tilde{L}_i^T\tilde{W}_i\{{Y}_i-g_\delta (\underline{X}_i)  - \tilde{m}_i(T)\}=0, \label{ee}
\end{align}
where 
\begin{align*}
    \tilde{L}_i = \Dot{g}_\delta(\underline{X}_i) - E^{-1}(\tilde{W}_i)E\left\{\tilde{W}_i\Dot{g}_\delta(\underline{X}_i)\right\},
\end{align*}
$\tilde{{W}}_i$ is the inverse of an arbitrary working correlation matrix, and $\tilde{m}_i(T)$ is a $\sum_{j}{n}_{ij}$ vector of arbitrary working control time trend functions. 

\begin{thm}
    The estimator arising from (\ref{ee}) is consistent and asymptotically normal, even when $\tilde{m}_i(T)$ and $\tilde{W}_i$ are misspecified. When $\tilde{W}_i$ and $\tilde{m}_i(T)$ are correctly specified, it is a semiparametric efficient estimator in the sense that no regular estimator under restriction (\ref{id_trteffect}) can have a smaller asymptotic variance.
    \label{thm2}
\end{thm}

\textcolor{black}{The theoretical properties of the estimator are nontrivial because the cluster-level contributions need not be identically distributed, and the restricted randomization induces dependence through the assignment mechanism. We therefore decompose the estimating equation into an outcome-noise term and a randomization-noise term. The outcome-noise term is handled using a Lindeberg-type central limit theorem for independent but non-identically distributed cluster-level contributions, whereas the randomization-noise term is handled using a combinatorial central limit theorem. Details are given in Section 3 of the Supplementary Materials.}

The robustness in Theorem \ref{thm2} arises from the construction of $\tilde{L}_i$ in (\ref{ee}). Under constraint (\ref{id_trteffect}), the conditional expectation $E\{Y_i - g_\delta(\underline{X}_i) - \tilde{m}_i(T) \mid \underline{X}_i\}$ does not depend on $\underline{X}_i$ for any $\tilde{m}_i$, so it is uncorrelated with $\tilde{L}_i$. Randomization further ensures that the distribution of $\underline{X}_i$ is known to be a permutation distribution, so the expectation in the definition of $\tilde{L}_i$ can be explicitly calculated, and the expectation of $\tilde{L}_i^T \tilde{W}_i$ is always zero. Therefore, the estimating equation (\ref{ee}) is always unbiased.

Such robustness does not hold in the commonly used linear mixed model (LMM) models like (\ref{lmmgen}) or generalized estimating equations (GEE) with an equivalent mean model, because they rely on the following estimating equation:
\begin{align}
    \sum_i \Dot{g}_\delta(\underline{X}_i)^T \tilde{W}_i \{{Y}_i - g_\delta(\underline{X}_i) - \tilde{m}_i(T)\} = 0. \label{lmmee}
\end{align}
Without using the centered term $\tilde{L}_i$ in (\ref{ee}), a GEE estimator based on (\ref{lmmee}) is biased unless $\tilde{m}_i(T)$ is correctly specified. Additionally, correctly specified random effects (\emph{i.e.}, $\tilde{W}_i$) are further required for the linear mixed model to provide valid inference based on likelihood methods.   

The choice of $\tilde{{W}_i}$ and $\tilde{m}_i(T)$ only affects the efficiency of the estimator from (\ref{ee}). Generally speaking, the closer $\tilde{W}_i$ and $\tilde{m}_i(T)$ are to their true values, the more efficient the estimator is. Since the nuisance parameter $\tilde{m}_i(T)$ captures the time-trend, it can be modeled categorically or as a continuous function of time. The choice of $\tilde{W}_i$ is similar to the choice of working correlation matrix in the general estimating equations framework. For example, when the correlation between any two observations in the same cluster is believed to be constant, an exchangeable working correlation structure can be used. When the control time trend is set to $0$, and the $\tilde{W}_i$ is set to be the identity matrix, the resulting estimator coincides with the robust estimator proposed in \cite{hughes2020robust}. However, this specification is likely suboptimal and will incur a notable loss in efficiency compared to a well-chosen $\tilde{W}_i$ and $\tilde{m}_i$.

\subsection{Inference}
Under a slightly restricted set of treatment effect models of the form $g_\delta (\underline{X}_i) = g (\underline{X}_i) \delta$, which covers both IT model (\ref{instant}) and ETI model(\ref{eti}), the estimating equation (\ref{ee}) takes the form
\begin{align*}
   \sum_i \tilde{L}^T_i \tilde{W}_i \left( Y_i - \tilde{m}_i(T) \right) -  \left\{\sum_i \tilde{L}^T_i \tilde{W}_i {g}(\underline{X}_i)\right\}\delta = 0,
\end{align*}
for which we obtain a closed-form estimator $\tilde{\delta}$ for $\delta$:
\begin{align}
    \tilde{\delta} = \left\{\sum_i \tilde{L}^T_i \tilde{W}_i {g}(\underline{X}_i)\right\}^{-1} \sum_i \tilde{L}^T_i \tilde{W}_i \left( Y_i - \tilde{m}_i(T) \right). \label{sameest}
\end{align}
Denote $\tilde{Y}_{0i;\delta} = Y_i - {g}(\underline{X}_i)\delta - \tilde{m}_i(T)$. We base inference on the exact variance of $\tilde{\delta}$, $\operatorname{Var}(\tilde{\delta})$, which is
\small
\begin{align}
   &E \left[ \left\{ \sum_i \tilde{L}_i^T \tilde{W}_i {g}(\underline{X}_i) \right\}^{-1} \left\{ \sum_i \tilde{L}_i^T \tilde{W}_i E\{\tilde{Y}_{0i;\delta} \tilde{Y}_{0i;\delta}^T\mid \underline{X}_i\} \tilde{W}_i \tilde{L}_i \right\} \left\{ \sum_i \tilde{L}_i^T \tilde{W}_i {g}(\underline{X}_i) \right\}^{-T} \right] \notag \\
    &+ E \left[ \left\{ \sum_i \tilde{L}_i^T \tilde{W}_i {g}(\underline{X}_i) \right\}^{-1} \left\{ \sum_{i \neq i'} \tilde{L}_i^T \tilde{W}_i E\{\tilde{Y}_{0i;\delta} \tilde{Y}_{0i';\delta}^T\mid \underline{X}_i, \underline{X}_{i'}\}\tilde{W}_{i'} \tilde{L}_{i'} \right\} \left\{ \sum_i \tilde{L}_i ^T\tilde{W}_i {g}(\underline{X}_i) \right\}^{-T} \right], \label{exactvariance}
\end{align}
\normalsize
where the outer expectations are taken with respect to the treatment assignment distribution, the inner expectations are taken with respect to the conditional distribution of outcome given treatment. For an invertible matrix $A$,  we denote $A^{-T}=(A^{-1})^{T}$. The exact variance $\operatorname{Var}(\tilde{\delta})$ contains both within cluster and cross cluster components, which are the first and second terms of (\ref{exactvariance}) respectively.  The former is analogous to the target of conventional sandwich variances and the latter arises because the permutation based assignment induces dependence across realized treatment paths and a possible misspecification of the working time trend model would yield nonzero cross cluster mean terms.
Motivated by this target, we propose a permutation variance estimator ${V}_\delta^e$, obtained by averaging the empirical analog of the exact variance formula over the assignment distribution:
\begin{align*}
 {V}_\delta^e= &E_p \left[ \left\{ \sum_i \tilde{L}_i^T \tilde{W}_i {g}(\underline{X}_i) \right\}^{-1} \left\{ \sum_i \tilde{L}_i^T \tilde{W}_i\tilde{Y}_{0i;\delta} \tilde{Y}_{0i;\delta}^T \tilde{W}_i \tilde{L}_i \right\} \left\{ \sum_i \tilde{L}_i^T \tilde{W}_i {g}(\underline{X}_i) \right\}^{-T} \right] \\
    &+ E_p \left[ \left\{ \sum_i \tilde{L}_i^T \tilde{W}_i {g}(\underline{X}_i) \right\}^{-1} \left\{ \sum_{i \neq i'} \tilde{L}_i^T \tilde{W}_i \tilde{Y}_{0i;\delta} \tilde{Y}_{0i';\delta}^T  \tilde{W}_{i'} \tilde{L}_{i'} \right\} \left\{ \sum_i \tilde{L}_i^T \tilde{W}_i {g}(\underline{X}_i) \right\}^{-T} \right],
\end{align*}
where $E_p$ is the expectation with respect to the permutation distribution of the treatment assignment.  
Under the same assumptions as Lemma 2, we show in Supplementary Material Section 4 that ${V}_\delta^e$ is consistent for $\operatorname{Var}(\tilde{\delta})$:
\begin{thm}
    Under the regularity conditions of Supplementary Lemma 2,  the proposed variance estimator $V_\delta^e$ is consistent for  $\operatorname{Var}(\tilde{\delta})$.
\end{thm}
In practice, when constructing $\tilde{Y}_{0i;\delta}$, we use a leave-one-cluster-out estimator $\tilde{\delta}_{-i},$ in place of $\delta$, where $\tilde{\delta}_{-i}$ is estimated using all clusters except cluster $i$, to reduce plug-in bias in the variance estimator. 
In addition, the estimating equation is Neyman orthogonal with respect to the nuisance estimators $\tilde{m}_i(T)$ and $\tilde{W}_i$, so nuisance estimation error does not affect the estimator at first order.  Under equal cluster sizes, $\tilde{\delta}$ is unbiased for $\delta$, and if $\tilde{W}_i$ is common across clusters, the proposed variance estimator is unbiased for the exact randomization variance. These additional properties are proved in Supplementary Material Section 5. 

\subsection{Handling Cluster-Level Covariates and Imbalances}
Proper handling of cluster-level covariates is important because (1) they may modify the treatment effect, thereby altering the form of the treatment effect model, and (2) they may remain imbalanced even after randomization, since it is typical for a stepped-wedge trial to have a small number of clusters, which in turn corresponds to a high likelihood of imbalance by chance.  
\subsubsection{Effect modification}
In the presence of potential cluster-level effect modifiers $S_{i}$ (e.g., baseline cluster sizes), it may be more appropriate to assume a treatment effect model that depends on $S_{i}$ to study effect modification:
\begin{align}
    E[Y_{ijk}(\bar{x}_{ij})|S_{i}] - E[Y_{ijk}(\bar{0}_j)|S_{i}] = g_\delta(\bar{x}_{ij}, S_{i}), \label{trteffect_general}
\end{align}
where $\delta$ typically includes main effects and interactions with $S_i$. 

The proposed method can be straightforwardly extended to settings that include effect modifiers. To see this, we examine the proposed estimating equation incorporating $S_i$:
\begin{align}
    \sum_i\tilde{L}_i^T\tilde{W}_i\{{Y}_i-g_\delta (\underline{X}_i,S_i)  - \tilde{m}(T,S_i)\}=0, 
    \label{general_ee}
\end{align}
where $\tilde{L}_i$ and $\tilde{W}_i$ are defined as in (\ref{ee}), and $\tilde{m}(T,S_i)$ is a $\sum_{ij}{n}_{ij}$ vector of arbitrary working functions of both time trend and $S_i$. Note that
\begin{align*}
    E\{{Y}_i-g_\delta (\underline{X}_i,S_i)  - \tilde{m}(T,S_i)|\underline{X}_i\}  = \int E\{{Y}_i-g_\delta (\underline{X}_i,S_i)  - \tilde{m}(T,S_i)|\underline{X}_i,S_i\}dF(S_i|\underline{X}_i)
\end{align*}
where $E\{{Y}_i-g_\delta (\underline{X}_i,S_i)  - \tilde{m}(T,S_i)|\underline{X}_i,S_i\}$ does not depend on $\underline{X}_i$ according to \eqref{trteffect_general}, and $F(S_i|\underline{X}_i)$, the conditional distribution of $S_i$ given $\underline{X}_i$, equals the marginal distribution of $S_i$ due to randomization. Therefore, the conditional expectation $E\{{Y}_i-g_\delta (\underline{X}_i,S_i)  - \tilde{m}(T,S_i)|\underline{X}_i\}$ does not depend on $\underline{X}_i$. Following the same argument as before, the zero expectation of $\tilde{L}_i^T\tilde{W}_i$ ensures that the estimating equation (\ref{general_ee}) is unbiased. 

The choice between assumptions (\ref{trteffect}) or (\ref{trteffect_general}) depends on which model better reflects scientific beliefs and interpretation of interest.

\subsubsection{Imbalanced precision variables}
In randomized experiments, any (vector of) covariates $K_i$ that is predictive of the outcome but not causally associated with the treatment does not act as a confounder, but as a precision variable. Randomization ensures that, on average, treatment assignment is independent of such covariates. Therefore, the presence of imbalance in $K_i$ should not affect the validity of the treatment effect estimates on average even if there is no adjustment for $K_i$.  However, in finite samples, randomization can result in imbalances by chance, leading to in-sample association between treatment and certain covariates.  When this happens, estimates may show a notable deviation from their true value when $K_i$ is not adjusted for. While valid on average, a sizable in-sample association can lead to misleading estimates in a particular trial. 

In the literature of randomized experiments, design-based \citep{morgan2012rerandomization, miratrix2013adjusting} and model-based adjustments \citep{fisher1971design, lin2013} have been advocated, and double adjustment combining the two has been studied \citep{li2020rerandomization}.  The proposed method for SWDs can seamlessly incorporate design and model-based adjustments for imbalances in $K_i$.  

For design-based adjustment, we could modify $\tilde{L}_i$ to include information from $K_i$ to improve precision. Specifically, we can use $\hat{E}^{-1}(W_i \mid K_i)\hat{E}\left\{ W_i \Dot{g}_\delta(\underline{X}_i)\mid K_i \right\}$  instead of ${E}^{-1}(W_i){E}\left\{ W_i \Dot{g}_\delta(\underline{X}_i) \right\}$ as the centering term in $\tilde{L}_i$, where $\hat{E}$ is estimated based on modeling or stratification using the observed $K_i$. Importantly, this design-based adjustment relies solely on the treatment and covariates, without using the outcome, which aligns with the foundational principles of propensity score methodology, which preserves the objectivity of the causal inference process, and strengthens the validity of the resulting estimates by avoiding data snooping \citep{rubin2007design}.

This design-based adjustment is closely connected to post-stratification approaches \citep{miratrix2013adjusting} for improving precision with random covariate imbalance. For a categorical variable $K$ with strata $\kappa \in \mathcal{K}$, our estimator can be viewed as a reweighted version of the post-stratification estimator, with weights that reflect both stratum size and within-stratum treatment allocation. To illustrate the connection, consider the instantaneous effect model with $T=3$, equal cluster size ($n_2$) at $j=2$, an independence working correlation matrix, $\tilde{m}_i(T)=0$, and a stratum-specific centering of $\tilde{L}_i$. In this case, the proposed estimator reduces to 
\begin{align*}
\tilde{\delta} = \sum_{\kappa \in \mathcal{K}} \dfrac{\frac{w_\kappa(0)w_\kappa(1)}{n_\kappa}}{\sum_{\kappa \in \mathcal{K}} \frac{w_\kappa(1)w_\kappa(0)}{n_\kappa}}\left\{\sum_{i \in S_\kappa}\frac{A_i\bar{Y}_i}{w_\kappa(1)} - \sum_{i \in S_\kappa}\frac{(1-A_i)\bar{Y}_i}{w_\kappa(0)}\right\},
\end{align*}
where $A_i = X_{i2}$, $\bar{Y}_i = \frac{1}{n_2}\sum_{k=1}^{n_2}Y_{i2k}$, $w_\kappa(a)=\sum_{i\in S_\kappa}\mathcal{I}(A_i=a)$, and $S_\kappa$ is the set of $n_\kappa$ clusters in stratum $\kappa$. See Supplementary Material Section 6. 

In comparison, the post-stratification estimator for cluster means at $T=2$ is \begin{align*}
    \hat{\delta}_{\text{PS}} = \sum_{\kappa \in \mathcal{K}}\frac{n_\kappa}{N}\left\{\sum_{i \in S_\kappa}\frac{A_i\bar{Y}_i}{w_\kappa(1)} - \sum_{i \in S_\kappa}\frac{(1-A_i)\bar{Y}_i}{w_\kappa(0)}\right\}.
\end{align*}
The weighting term in $\hat{\delta}_{PS}$ is purely based on the stratum size, whereas the weighting in $\tilde{\delta}$ can be interpreted as an inverse variance weight, since the variance of the within-stratum average treatment effect estimator is inversely proportional to this quantity under constant outcome variance. 

In contrast to design-based adjustments, model-based adjustments modify the working control cluster-period mean, $\tilde{m}_i$, by explicitly including $K_i$, \emph{i.e.} we use a working model $\tilde{m}_i(T,K_i)$.  Including baseline covariates in the working baseline mean model is conceptually similar to robust regression adjustments such as ANCOVA. Our proposed method offers opportunities for ``double adjustment'' using both the design-based adjustment in $\tilde{L}_i$ and the model-based adjustment in $\tilde{m}_i$. 

\section{Simulation}
We present two sets of simulation studies; one demonstrates the robustness of the proposed estimator under a misspecified time trend and variance-covariance structure and efficiency under correctly specified models, the other one demonstrates the performance of the proposed estimator under covariate imbalance, which typically occurs when the number of clusters is small. For both simulation studies, we compare the proposed estimator with the commonly used linear mixed model. For inference, we compare the coverage of the $95\%$ confidence intervals using the model-based standard error estimator for LMMs, a conventional sandwich estimator and the proposed permutation-based standard error estimators with and without leave-one-out correction. 

\subsection{Setting 1: Robustness under misspecifications}
We consider a cross-sectional SWD with 10 clusters and 5 time points, where there are two or three clusters per sequence. Sequences can have unequal numbers of clusters, and cluster-to-sequence assignments are re-randomized for each experiment. The cluster-period sizes are not restricted to be the same within or across clusters. They were generated according to four patterns, with baseline sizes between 11 and 20. In two of the patterns, cluster sizes remained constant over time, while in the other two, they increased by one at every time point.
The outcome is generated as follows:
\begin{align*}
    Y_{ijk} = 3 + 4(-j + 1)^2 + 4x_{ij} + \tau_i + \eta_{i} \cdot j + \epsilon_{ijk},
\end{align*}
where $x_{ij}$ is the indicator for treatment for cluster $i$ at time $j$, $\tau_i$ is a random intercept capturing the random cluster effect, and $\eta_{i}$ is a random slope for continuous time.  The individual error term $\epsilon_{ijk}$, the random intercept $\tau_i$, and the random slope $\eta_{ij}$ are mutually independent, each with a mean of $0$. The variances of $\tau_i$, $\eta_{ij}$, and $\epsilon_{ijk}$ are $0.25$, $0.25$, and $4$. 

The scenarios and estimators under comparison are shown in Table \ref{tab:model_comparison1}. Models labeled as `correct' accurately represent the corresponding aspect of the data generating mechanism.
We consider various fixed and random effect combinations for both LMM and the proposed estimator, where fixed and random effects are referred to the working control cluster-period mean and working covariance models in the proposed method.
\begin{table}
\caption{\label{tab:model_comparison1} Model comparison table for Setting 1}
\centering
\begin{tabular}{c|c|c|c}
\hline
\textbf{Model} & \multicolumn{2}{c|}{\textbf{Model Specification}} & \textbf{Estimator} \\ 
\cline{2-3}
               & \textbf{Fixed Effect} & \textbf{Random Effect} & \\ 
\hline
(a)            & Linear 
               & cluster effect 
               & LMM-(a)\\ 
               & (incorrect) 
               & (incorrect) 
               & Proposed-(a) \\ 
\hline
(b)            & Linear
               & cluster and time effect 
               & LMM-(b)\\ 
               & (incorrect) 
               & (correct) 
               & Proposed-(b) \\ 
\hline
(c)            & Categorical
               & cluster effect 
               & LMM-(c)\\ 
               & (correct) 
               & (incorrect) 
               & Proposed-(c) \\ 
\hline
(d)            & Categorical
               & cluster and time effect
               & LMM-(d)\\ 
               & (correct) 
               & (correct) 
               & Proposed-(d) \\ 
\hline
\end{tabular}%
\end{table}

\begin{table}
\small
\caption{\label{tab:res1} Simulation results for Setting 1 based on 1000 experiments. S.E. represents the standard error of the estimator across all experiments. Model Cov. and Sand. Cov. refer to the coverage of the 95\% model-based and sandwich-based confidence interval for LMM. Perm. Cov. and Perm-L1O Cov. indicate the coverage of the 95\% confidence intervals obtained using the proposed permutation-based standard error estimator without and with leave-one-out correction, respectively. Fail counts the number of experiments in which the method failed to fit out of 1000.}
\centering
\begin{tabular}{c|c|c|c|c|c|c|c}
\hline
Model & Bias & S.E. & Model & Sand. & Perm. & Perm-L1O & Fail \\
      &      &      & Cov. (\%) & Cov. (\%) & Cov. (\%) & Cov. (\%) & \\
\hline
LMM-(a)      & 0.05 & 1.05 & 86\% & --   & --    & --    & 0  \\
Proposed-(a) & 0.01 & 0.36 & --    & 100\% & 87\% & 98\%  & -- \\
LMM-(b)      & 0.05 & 0.96 & 88\% & --   & --    & --    & 30 \\
Proposed-(b) & 0.00 & 0.31 & --    & 100\% & 87\% & 98\%  & -- \\
LMM-(c)      & 0.01 & 0.32 & 88\% & --   & --    & --    & 0  \\
Proposed-(c) & 0.01 & 0.33 & --    & 86\% & 91\%  & 95\%  & -- \\
LMM-(d)      & 0.01 & 0.26 & 92\% & --   & --    & --    & 38 \\
Proposed-(d) & 0.00 & 0.26 & --    & 86\% & 90\%  & 94\%  & -- \\
\hline
\end{tabular}
\end{table}


As shown in Table \ref{tab:res1}, the LMM estimators exhibit some bias when the time trend is misspecified, and the model-based coverage notably falls below the nominal 95\% when the random effect is misspecified.  The proposed estimators exhibit negligible bias regardless of the time-trend specification. 
\textcolor{black}{When the control time trend is misspecified, the sandwich variance estimator tends to overestimate the variability because it ignores a non-positive cross term in the exact variance. By contrast, when the control time trend is correctly specified, the sandwich estimator tends to underestimate the standard error as seen in conventional GEE when the number of clusters is small.
The permutation based standard error estimator without leave-one-out correction performs better than the sandwich estimator, but still tends to undercover. This undercoverage may arise partly from plug-in error due to the estimator $\hat{\delta}$, and partly from the use of normal approximations to construct Wald confidence intervals when the number of clusters is small. The permutation estimator with leave-one-out correction leads to a notable improvement in coverage.}

When the LMM models are correctly specified, the performances of LMM and proposed estimators are comparable. This indicates that the proposed semiparametric estimator has a minimal loss in efficiency compared to the parametric estimator when the model is appropriately specified. Table \ref{tab:res1} includes a column labeled \textit{Fail}, which reports the number of experiments in which the model failed to fit using the R package \textsc{glmmTMB}.

\subsection{Setting 2: Stability under covariate imbalance}
Next, we consider a conventional cross-sectional SWD with the same number of clusters, time periods and cluster-period size distribution as in Setting 1, but a different outcome model with a non-linear interaction between baseline cluster size and time trend:
\begin{align*}
 Y_{ijk} = 3 +  I(K_{i1} > 15)4(-j + 1)^2 + 4x_{ij} + \tau_i + \eta_{i} \cdot j + \epsilon_{ijk},   
\end{align*} 
where $K_{i1}$ is the baseline cluster size for cluster $i$, and $I(K_{i1} \geq 15)$ is an indicator of whether the baseline cluster size is greater than or equal to 15. In this model, only these clusters have a time trend that is nonlinear. 
The distributions of $\tau_i$, $\eta_{ij}$, and $\epsilon_{ijk}$ are the same as in Setting 1.

As explained in Section 2.4.2, since $K_{i1}$ is independent of treatment assignment, the terms involving $K_{i1}$ will average across its distribution when considering the expectation of the outcome given the treatment. Consequently, a model specification that correctly accounts for confounding by time will yield unbiased point estimates for the treatment effect on average, regardless of whether the dependency between outcome and $K_{i1}$ is accurately modeled. However, with a small number of clusters, $K_{i1}$ can have in-sample positive or negative correlation with treatment by chance, leading to estimates for an individual study that deviate from the true value.

The scenarios and estimators under comparison are shown in Table \ref{tab:model_comparison2}. Models labeled as `correct' accurately represent the data generating mechanism. We evaluate various combinations of fixed and random effects for both LMM and proposed methods. Baseline cluster size is included as a continuous main effect (incorrectly) in the outcome model for both LMM and the proposed estimator, reflecting a simple form of adjustment that may be adopted in practice. For the proposed estimator, we further employ the design-based adjustment of $K_{i1}$ as discussed in Section 2.4.2.

\begin{table}
\caption{\label{tab:model_comparison2} Model comparison table for Setting 2}
\centering
\begin{tabular}{c|c|c|c|c}
\hline
 & \multicolumn{3}{c|}{\textbf{Model Specification}} & \\ 
\cline{2-4}
 \textbf{Model}              & \multicolumn{2}{c|}{\textbf{Fixed Effect}} & \textbf{Random Effect} & \textbf{Estimator}\\ 
\cline{2-3}
               & Time Trend & Precision variable & & \\ 
               &&(incorrect)&&\\
\hline
(e)            & Linear & Baseline 
               & Cluster effect 
               & LMM-(e)\\
               &(incorrect) & cluster size &(incorrect)  &Proposed-(e) \\
\hline
(f)            & Linear & Baseline
               & Cluster and time effect
               & LMM-(f) \\ 
               &(incorrect) & cluster size &(correct)  &Proposed-(f)  \\
\hline
(g)            & Categorical  & Baseline
               & Cluster effect
               & LMM-(g)\\ 
               &(correct) & cluster size &(incorrect)  &Proposed-(g)  \\
\hline
(h)            & Categorical& Baseline
               & Cluster and time effect
               & LMM-(h)\\
               &(correct) & cluster size &(correct)  &Proposed-(h)  \\
\hline
\end{tabular}
\end{table}

\begin{table}
\caption{\label{tab:res2} Simulation results for Setting 2. S.E. represents the standard error of the estimator across all experiments. Model Cov. refers to the coverage of the 95\% model-based confidence interval for LMM. Perm. Cov. and Perm-L1O Cov. indicate the coverage of the 95\% confidence intervals obtained using the proposed permutation-based standard error estimator without and with leave-one-out correction, respectively. Fail counts the number of experiments in which the method failed to fit out of 1000. }
\centering
\small
\begin{tabular}{c|c|c|c|c|c|c|c}
  \hline
  {Model} & {Bias} & {S.E.} & {Model}  &Sand. & {Perm.} & {Perm-L1O} & {Fail} \\ 
  & & & {Cov.(\%)}  &Cov.(\%) & {Cov.(\%)}  & {Cov.(\%)} & \\ 
  \hline
  LMM-(e) & -0.03 & 4.48  & 50\%  &--& -- &--  & 0 \\ 
  \textit{Proposed-(e)} & 0.00 & 0.43  & --  &100\%& 90\% & 95\% & 0 \\ 
  LMM-(f) & 0.09 & 3.01  & 33\% &--& -- &--  & 35\\ 
  \textit{Proposed-(f)} & -0.00 & 0.45  & --  &100\%& 88\% & 95\% & 0 \\ 
  LMM-(g) & -0.07 & 4.02 & 52\%  &--&--  & -- & 0\\ 
  \textit{Proposed-(g)} & 0.00 & 0.43  & --  &100\%& 90\% & 95\% & 0 \\ 
  LMM-(h) & 0.07 & 2.53 & 20\%  &--&--  &--  & 8 \\ 
  \textit{Proposed-(h)} & 0.01 & 0.35  & --  &100\%& 88\% & 96\% & 0 \\ 
   \hline
\end{tabular}
\end{table}

\begin{figure}[h!]
    \centering
    \includegraphics[width=0.8\textwidth]{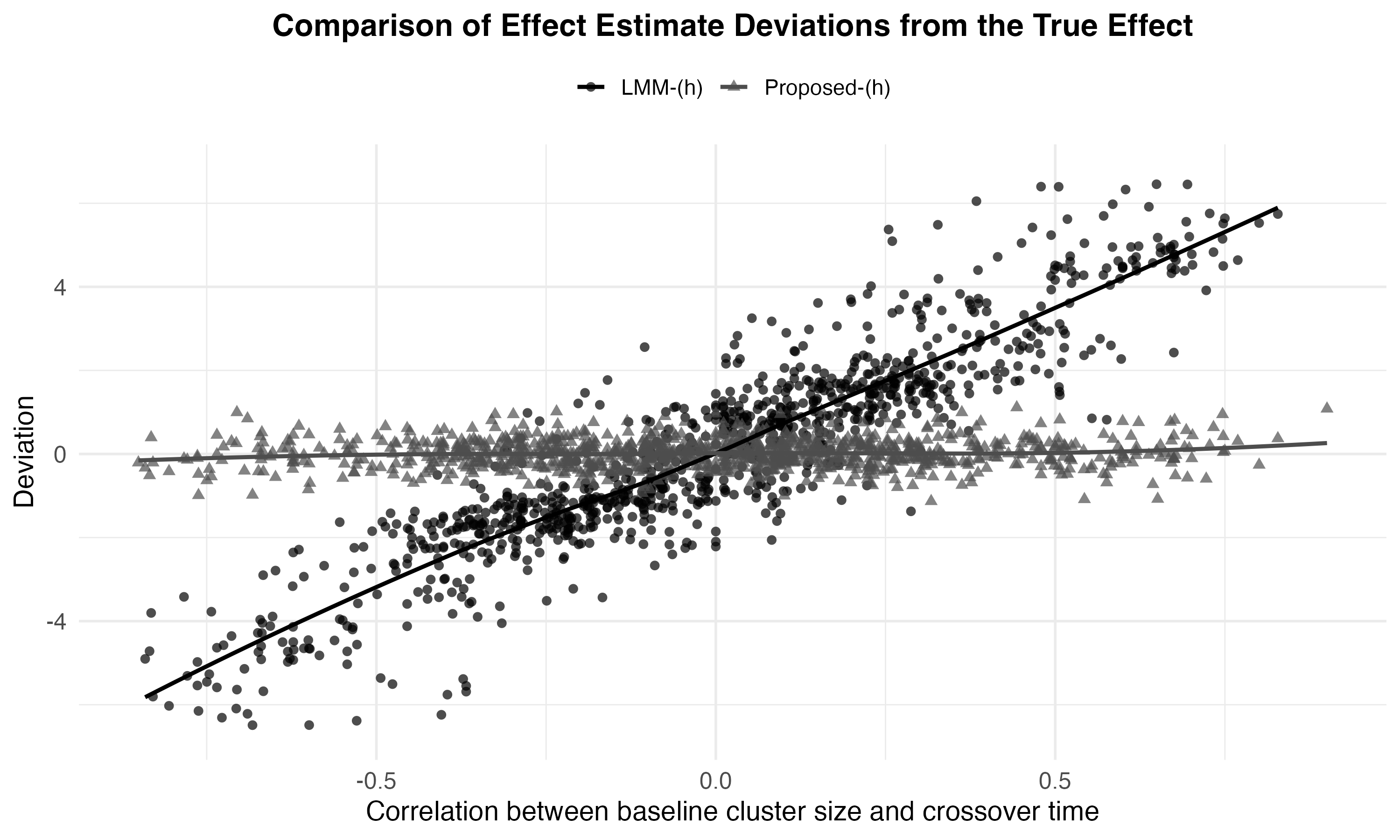} 
     \caption{Comparison between LMM-(h) (black circles) and Proposed-(h) (grey triangles). Each point represents the difference between the point estimate and the true treatment effect for a single experiment, with a total of 1000 experiments. Loess curves are included to depict trend.}
    \label{fig:deviation}
\end{figure}


%

As shown in Table \ref{tab:res2}, the model-based coverage of LMM estimators often falls substantially below the nominal 95\% due to misspecification of random effects or heteroskedasticity, which arises from failing to account for the interaction between the imbalanced covariate and time. The presence of non-linear interactions requires the inclusion of terms like squared time in the random effects model for accurate specification, which is seldom employed in practice. While configuring an LMM with a highly flexible random effects structure can potentially address this under-coverage, such complexity could cause model non-convergence particularly with small cluster sizes. 


The proposed estimators exhibit notably smaller deviations from the truth compared to LMM in Table \ref{tab:res2}, indicating that the design-based adjustment can notably reduce the estimation error due to the presence of in-sample correlation by chance when the model-based adjustment fails. 

In Figure \ref{fig:deviation}, we further demonstrate the difference between LMM and the proposed method under Model (h).   Similar comparisons for Models (e), (f) and (g) are provided in Section 7 of Supplementary Materials.  The deviation of LMM estimators from the truth is linearly related to the in-sample correlation between the potentially imbalanced baseline cluster size and the treatment crossover time, resulting in a large standard error.  The deviation of the proposed estimator from the truth does not appear to vary with the in-sample correlation between the imbalanced covariate and the treatment assignment. This results in a more stable estimate for a small number of clusters compared to LMM. As the number of clusters increases, the deviation between in-sample correlation and the true value $0$ decreases. Consequently, the deviation between LMM and the proposed estimator also decreases with an increase in the number of clusters. These results are included in Section 7 of Supplementary Materials. However, a small number of clusters is the rule rather than the exception in practical implementation of stepped wedge trials, and therefore, the issue with imbalanced baseline variables often needs to be adequately addressed.

\section{Real Data Application: Washington State expedited partner treatment trial}
We use data from the Washington State Community-Level Expedited Partner Treatment (EPT) Randomized Trial \citep{golden2015uptake} to illustrate our method and compare it with conventional mixed models. The trial evaluated whether EPT, which treats the partners of individuals with sexually transmitted infections without requiring a medical evaluation, reduces chlamydia and gonorrhea rates. The dataset includes 167,989 individuals across 22 clusters (local health jurisdictions), 4 sequences, and 5 time points. We excluded 6,830 individuals (approximately $4\%$) with missing values in variables required for the analysis, including cluster ID, time, treatment assignment, and outcome, resulting in a final analytic sample of 161,159 individuals. The effect of interest is the difference in chlamydia rates.


We observe substantial variation in baseline cluster size, ranging from 30 to 991 individuals. Moreover, baseline cluster size is moderately correlated with crossover time (Pearson’s correlation coefficient = –0.31), suggesting that it may be imbalanced. Importantly, we do not use outcome data to determine whether or how baseline cluster size should be adjusted for in the analysis. 

\textcolor{black}{We apply the proposed method to the data under an instantaneous treatment (IT) effect model. The working model includes categorical time effects and uses an exchangeable working correlation structure. We compare the resulting estimates to those from two linear mixed models, both of which include categorical time and log baseline cluster size as fixed effects but differ in their random effect structures: (1) a random intercept for cluster (LMM-1), and (2) a random intercept for cluster along with a random time effect nested within cluster (LMM-2). For the mixed models, we report both model-based confidence intervals and percentile bootstrap confidence intervals. The bootstrap confidence intervals were computed from the 2.5th and 97.5th percentiles of the bootstrap distribution. (An analysis not adjusted for baseline cluster size is presented in Section 8 of the Supplementary Material.)
}
\begin{table}
\caption{\label{rara2}Estimated treatment effect of EPT, adjusted for log baseline cluster size}
\centering
\begin{tabular}{c|c|c|c|c}
\hline
\textbf{Model} & \textbf{Estimate} & \textbf{SE Method} & \textbf{SE} & \textbf{$95\%$ CI} \\
\hline
Proposed Method & $-0.0086$ & Permutation-based SE &$0.0045$ &$(-0.0174, 0.0002)$ \\
                &           & Leave-one-out SE     & $0.0047$ &$(-0.0178,0.0005)$ \\
\hline
LMM-1 & $-0.0087$ & model-based SE & $0.0024$ &$(-0.0134, -0.0039)$ \\
LMM-2 &  $-0.0014$& model-based SE & $0.0037$ & $(-0.0086, 0.0058)$ \\
\hline
\end{tabular}
\end{table}

From Table \ref{rara2}, we observe that the point estimates from the proposed method are similar to those from the linear mixed model with a random intercept only. This is expected, as the estimating equation used by the proposed method is similar to that of LMM-1. However, the standard errors estimated by the proposed method are larger than those from both mixed models, suggesting that the model-based standard errors may be anti-conservative. Notably, the 95\% confidence intervals from the two linear mixed models do not include each other’s point estimates. This inconsistency arises solely from differences in random effect specification, and suggests that at least one of the mixed models could be misspecified, rendering the associated model-based inference unreliable.  
\textcolor{black}{
We believe neither mixed model captures the full covariance structure adequately, because observations are additionally clustered within sites nested inside local health jurisdictions. As a result, the model-based standard errors for the LMMs are likely anti-conservative. In contrast, the $95\%$ confidence intervals constructed using the proposed method contain both point estimates from the two mixed models, which is expected under correct specification of the mean structure.  Our result is also consistent with the bootstrap confidence intervals for both LMM specifications: LMM-1 (-0.0156, -0.0022) and LMM-2 (-0.0129, 0.0072), which are notably wider than the corresponding model-based LMM intervals.}

\begin{figure}[ht]
\centering
\includegraphics[width=0.8\textwidth]{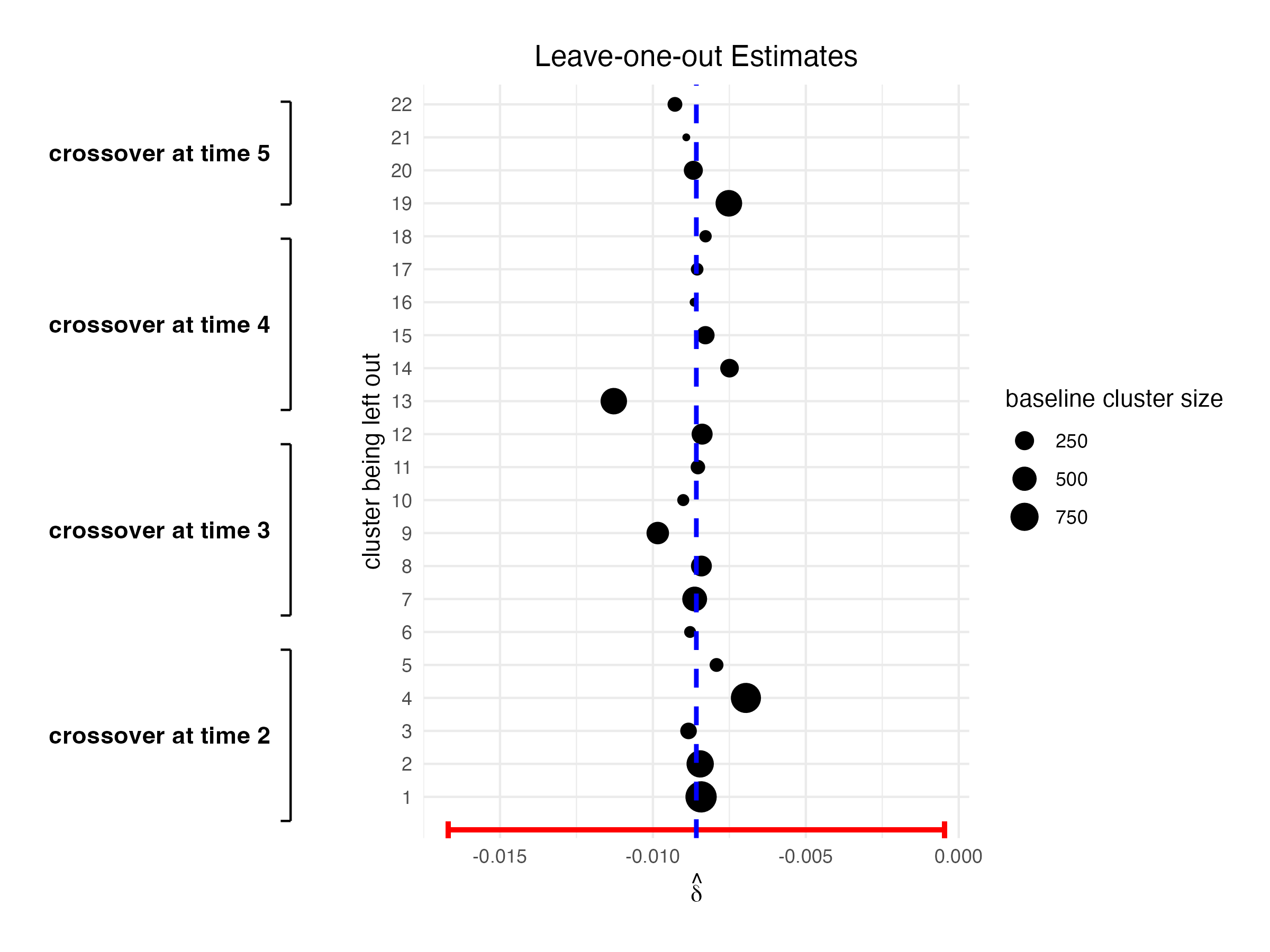}
\caption{Estimates of the treatment effect when leaving out each of the clusters.  The dashed line indicates the treatment effect estimate when using all clusters; dots further from the overall treatment effect indicate greater influence of the cluster being left out.  Leave-one-out estimates are ordered by crossover time; dot size reflects baseline cluster size.}
\label{fig:loo_plot_EPT}
\end{figure}

The leave-one-out standard error estimates are slightly larger than the permutation-based estimates, consistent with their conservative theoretical property. The small difference between the two suggests that few clusters exert substantial influence on the effect estimates, as illustrated in Figure~\ref{fig:loo_plot_EPT}.

\section{Discussion}
In this paper, we propose a unified framework for semiparametric effect modeling, estimation and inference for stepped wedge trials. Our estimator is flexible across a range of treatment effect models and is robust to misspecification of the baseline time trend and variance–covariance structure, which are issues known to be challenging in the existing literature. For a meaningful subclass of treatment effect models, we further propose a permutation-based standard error estimator to quantify finite-sample uncertainty, and we enhance its practical performance using a leave-one-out plug-in estimate.  We develop a semiparametric efficiency theory for stepped-wedge trials in the general setting of unequal cluster-period sizes and permutation-induced dependence. In contrast to earlier analyses, we neither assume equal cluster-period sizes nor rely on cluster-period means defined under constant sizes across periods and treated as independent and identically distributed. Our framework instead treats variability in cluster-period sizes and the permutation of treatment assignments as fundamental features, rather than secondary complications, in the design and analysis.

While robustness is achieved through centering in $\tilde{L}$, doing so may lead to a cost in efficiency. When certain elements of the centered term $\tilde{L}_i$ are zero, the corresponding outcomes do not contribute to the estimation of $\delta$ because their contribution to the estimating equation (\ref{ee}) is zero regardless of the value of $\delta$. When the number of sequences is $T-1$, $W$ is not treatment dependent and under the IT model, the centering terms are zero whenever all clusters are simultaneously untreated $(j=1)$ or treated $(j=T)$, and data from those periods cannot inform estimation of $\delta$. 
Nonparametric estimators based solely on within-period comparisons also exclude such periods. However, unlike our estimator, they do not leverage within-cluster comparisons and are therefore less efficient.  On the other hand, generalized estimating equations and linear mixed models incorporate information from these periods but their validity relies on correct specification of the control time trend. In our simulation study with $5$ time points, the loss in efficiency comparing the semiparametric estimator and the parametric estimator is negligible when nuisance functions are correctly specified.  This is likely due to the fact that additional time-specific parameters are estimated in the parametric LMMs. 

A key feature of our method is the separation of treatment assignment modeling from outcome modeling. This decoupling allows us to adjust for potential imbalances present in the data by centering the treatment assignment using its estimated conditional expectation given baseline variables. In particular, by accounting for imbalanced precision variables, the standard error of the estimator can be reduced. For discrete cluster-level covariates, we show that this adjustment is closely related to post-stratification weighting. Moreover, our method allows both design-based and outcome-based adjustments, offering opportunities for ``double adjustment''. 

Although our method is motivated by stepped wedge trials, its utility extends beyond this framework. For instance, it can be easily adapted to observational stepped wedge designs with confounding as well as to observational longitudinal studies with staggered entry or treatment adoption.
Although we focus on continuous outcomes under linear models, the method readily extends to other collapsible link functions, such as the log link, thereby accommodating a broader range of outcome types. For non-collapsible links, such as the logit link for binary outcomes, reparameterization strategies such as those in \cite{richardson2017modeling} may allow possible extensions, which we plan to investigate in future work. 

\bibliographystyle{plainnat}  
\bibliography{reference.bib}

@article{morgan2012rerandomization,
  title={Rerandomization to improve covariate balance in experiments},
  author={Morgan, Kari Lock and Rubin, Donald B},
  journal={The Annals of Statistics},
  volume={40},
  number={2},
  pages={1263--1282},
  year={2012}
}

@article{miratrix2013adjusting,
  title={Adjusting treatment effect estimates by post-stratification in randomized experiments},
  author={Miratrix, Luke W and Sekhon, Jasjeet S and Yu, Bin},
  journal={Journal of the Royal Statistical Society Series B: Statistical Methodology},
  volume={75},
  number={2},
  pages={369--396},
  year={2013},
  publisher={Oxford University Press}
}

@article{hughes2020robust,
  title={Robust inference for the stepped wedge design},
  author={Hughes, James P and Heagerty, Patrick J and Xia, Fan and Ren, Yuqi},
  journal={Biometrics},
  volume={76},
  number={1},
  pages={119--130},
  year={2020},
  publisher={Wiley Online Library}
}

@book{diggle2002analysis,
	title = {Analysis of {Longitudinal} {Data}},
	isbn = {978-0-19-852484-7},
	publisher = {Oxford University Press},
	author = {Diggle, Peter J and Heagerty, Patrick J and Liang, Kung-yee. and Zeger, Scott L},
	month = jun,
	year = {2002}}

@article{hooper2016sample,
  title={Sample size calculation for stepped wedge and other longitudinal cluster randomised trials},
  author={Hooper, Richard and Teerenstra, Steven and de Hoop, Esther and Eldridge, Sandra},
  journal={Statistics in medicine},
  volume={35},
  number={26},
  pages={4718--4728},
  year={2016},
  publisher={Wiley Online Library}
}

@article{hussey2007design,
	title={Design and analysis of stepped wedge cluster randomized trials},
	author={Hussey, Michael A and Hughes, James P},
	journal={Contemporary clinical trials},
	volume={28},
	number={2},
	pages={182--191},
	year={2007},
	publisher={Elsevier}
}

@article{mdege2011systematic,
	title={Systematic review of stepped wedge cluster randomized trials shows that design is particularly used to evaluate interventions during routine implementation},
	author={Mdege, Noreen D and Man, Mei-See and Taylor, Celia A and Torgerson, David J},
	journal={Journal of clinical epidemiology},
	volume={64},
	number={9},
	pages={936--948},
	year={2011},
	publisher={Elsevier}
}

@article{voldal2022misspec,
author = {Voldal, Emily C. and Xia, Fan and Kenny, Avi and Heagerty, Patrick J. and Hughes, James P.},
title = {Model misspecification in stepped wedge trials: Random effects for time or treatment},
journal = {Statistics in Medicine},
volume = {41},
number = {10},
pages = {1751-1766},
year = {2022}
}

@article{hemming2015stepped,
  title={Stepped-wedge cluster randomised controlled trials: a generic framework including parallel and multiple-level designs},
  author={Hemming, Karla and Lilford, Richard and Girling, Alan J},
  journal={Statistics in medicine},
  volume={34},
  number={2},
  pages={181--196},
  year={2015},
  publisher={Wiley Online Library}
}

@article{ji2017randomization,
  title={Randomization inference for stepped-wedge cluster-randomized trials: an application to community-based health insurance},
  author={Ji, Xinyao and Fink, Gunther and Robyn, Paul Jacob and Small, Dylan S},
  journal={The Annals of Applied Statistics},
  volume={11},
  number={1},
  pages={1--20},
  year={2017},
  publisher={Institute of Mathematical Statistics}
}

@article{scott2017finite,
  title={Finite-sample corrected generalized estimating equation of population average treatment effects in stepped wedge cluster randomized trials},
  author={Scott, JoAnna M and deCamp, Allan and Juraska, Michal and Fay, Michael P and Gilbert, Peter B},
  journal={Statistical methods in medical research},
  volume={26},
  number={2},
  pages={583--597},
  year={2017},
  publisher={SAGE Publications Sage UK: London, England}
}

@article{thompson2018robust,
  title={Robust analysis of stepped wedge trials using cluster-level summaries within periods},
  author={Thompson, JA and Davey, C and Fielding, K and Hargreaves, JR and Hayes, RJ},
  journal={Statistics in medicine},
  volume={37},
  number={16},
  pages={2487--2500},
  year={2018},
  publisher={Wiley Online Library}
}

@article{thompson2017bias,
  title={Bias and inference from misspecified mixed-effect models in stepped wedge trial analysis},
  author={Thompson, Jennifer A and Fielding, Katherine L and Davey, Calum and Aiken, Alexander M and Hargreaves, James R and Hayes, Richard J},
  journal={Statistics in medicine},
  volume={36},
  number={23},
  pages={3670--3682},
  year={2017},
  publisher={Wiley Online Library}
}

@article{wang2017use,
  title={The use of permutation tests for the analysis of parallel and stepped-wedge cluster-randomized trials},
  author={Wang, Rui and De Gruttola, Victor},
  journal={Statistics in medicine},
  volume={36},
  number={18},
  pages={2831--2843},
  year={2017},
  publisher={Wiley Online Library}
}

@article{kennedy2020novel,
  title={Novel methods for the analysis of stepped wedge cluster randomized trials},
  author={Kennedy-Shaffer, Lee and De Gruttola, Victor and Lipsitch, Marc},
  journal={Statistics in medicine},
  volume={39},
  number={7},
  pages={815--844},
  year={2020},
  publisher={Wiley Online Library}
}

@article{kenny2022analysis,
  title={Analysis of stepped wedge cluster randomized trials in the presence of a time-varying treatment effect},
  author={Kenny, Avi and Voldal, Emily C and Xia, Fan and Heagerty, Patrick J and Hughes, James P},
  journal={Statistics in Medicine},
  volume={41},
  number={22},
  pages={4311--4339},
  year={2022},
  publisher={Wiley Online Library}
}

@article{kahan2023informative,
  title={Informative cluster size in cluster-randomised trials: A case study from the TRIGGER trial},
  author={Kahan, Brennan C and Li, Fan and Blette, Bryan and Jairath, Vipul and Copas, Andrew and Harhay, Michael},
  journal={Clinical Trials},
  volume={20},
  number={6},
  pages={661--669},
  year={2023},
  publisher={SAGE Publications Sage UK: London, England}
}

@article{richardson2017modeling,
  title={On modeling and estimation for the relative risk and risk difference},
  author={Richardson, Thomas S and Robins, James M and Wang, Linbo},
  journal={Journal of the American Statistical Association},
  volume={112},
  number={519},
  pages={1121--1130},
  year={2017},
  publisher={Taylor \& Francis}
}

@article{ouyang2024maintaining,
  title={Maintaining the validity of inference from linear mixed models in stepped-wedge cluster randomized trials under misspecified random-effects structures},
  author={Ouyang, Yongdong and Taljaard, Monica and Forbes, Andrew B and Li, Fan},
  journal={Statistical Methods in Medical Research},
  volume={33},
  number={9},
  pages={1497--1516},
  year={2024},
  publisher={SAGE Publications Sage UK: London, England}
}

@book{hernan2020causal,
  title        = {Causal Inference: What If},
  author       = {Hernán, Miguel A. and Robins, James M.},
  year         = {2020},
  publisher    = {Boca Raton: Chapman \& Hall/CRC}}

@book{fisher1971design,
  title={The design of experiments},
  author={Fisher, Ronald Aylmer},
  year={1971},
  publisher={Springer}
}

@article{lin2013,
author = {Winston Lin},
title = {{Agnostic notes on regression adjustments to experimental data: Reexamining Freedman’s critique}},
volume = {7},
journal = {The Annals of Applied Statistics},
number = {1},
publisher = {Institute of Mathematical Statistics},
pages = {295 -- 318},
year = {2013}}

@article{li2020rerandomization,
  title={Rerandomization and regression adjustment},
  author={Li, Xinran and Ding, Peng},
  journal={Journal of the Royal Statistical Society Series B: Statistical Methodology},
  volume={82},
  number={1},
  pages={241--268},
  year={2020},
  publisher={Oxford University Press}
}

@article{rubin2007design,
  title={The design versus the analysis of observational studies for causal effects: parallels with the design of randomized trials},
  author={Rubin, Donald B},
  journal={Statistics in medicine},
  volume={26},
  number={1},
  pages={20--36},
  year={2007},
  publisher={Wiley Online Library}
}

@article{golden2015uptake,
  title={Uptake and population-level impact of expedited partner therapy (EPT) on Chlamydia trachomatis and Neisseria gonorrhoeae: the Washington State community-level randomized trial of EPT},
  author={Golden, Matthew R and Kerani, Roxanne P and Stenger, Mark and Hughes, James P and Aubin, Mark and Malinski, Cheryl and Holmes, King K},
  journal={PLOS medicine},
  volume={12},
  number={1},
  pages={e1001777},
  year={2015},
  publisher={Public Library of Science San Francisco, CA USA}
}

@article{jarvik2020effect,
  title={The effect of including benchmark prevalence data of common imaging findings in spine image reports on health care utilization among adults undergoing spine imaging: a stepped-wedge randomized clinical trial},
  author={Jarvik, Jeffrey G and Meier, Eric N and James, Kathryn T and Gold, Laura S and Tan, Katherine W and Kessler, Larry G and Suri, Pradeep and Kallmes, David F and Cherkin, Daniel C and Deyo, Richard A and others},
  journal={JAMA network open},
  volume={3},
  number={9},
  pages={e2015713--e2015713},
  year={2020},
  publisher={American Medical Association}
}

@article{kasza2019impact,
  title={Impact of non-uniform correlation structure on sample size and power in multiple-period cluster randomised trials},
  author={Kasza, J and Hemming, K and Hooper, R and Matthews, JNS and Forbes, AB},
  journal={Statistical methods in medical research},
  volume={28},
  number={3},
  pages={703--716},
  year={2019},
  publisher={SAGE Publications Sage UK: London, England}
}

@article{billot2024should,
  title={How should a cluster randomized trial be analyzed?},
  author={Billot, Laurent and Copas, Andrew and Leyrat, Clemence and Forbes, Andrew and Turner, Elizabeth L},
  journal={Journal of epidemiology and population health},
  volume={72},
  number={1},
  pages={202196},
  year={2024},
  publisher={Elsevier}
}
\end{document}